\documentclass[aps,superscriptaddress,reprint,nofootinbib]{revtex4-1}

\usepackage{xcolor}
\usepackage{graphicx}
\usepackage{float}
\usepackage{mwe}
\usepackage{indentfirst}
\usepackage{algorithm}
\usepackage{algpseudocode}
\usepackage{amsfonts, amsmath, amssymb}
\usepackage{mathtools}
\usepackage{xurl}

\begin{document}
\title{\LARGE Massively Parallel Tensor Network State Algorithms on \\ Hybrid CPU-GPU Based Architectures}

\author{Andor Menczer}%
\affiliation{%
Strongly Correlated Systems "Lendület" Research Group,
Wigner Research Centre for Physics, H-1525, Budapest, Hungary
}%
\affiliation{%
Eötvös Loránd University, Budapest, Hungary
}%

\author{Örs Legeza}
\affiliation{%
Strongly Correlated Systems "Lendület" Research Group,
Wigner Research Centre for Physics, H-1525, Budapest, Hungary
}%
\affiliation{
Institute for Advanced Study,Technical University of Munich, Lichtenbergstrasse 2a, 85748 Garching, Germany
}

\date{\today}

\begin{abstract}
	\noindent The interplay of quantum and classical simulation and the delicate divide between them is in the focus of massively parallelized tensor network state (TNS) algorithms designed for high performance computing (HPC). In this contribution, we present novel algorithmic solutions together with implementation details to extend current limits of TNS algorithms on HPC infrastructure building on state-of-the-art hardware and software technologies. Benchmark results obtained via large-scale density matrix renormalization group (DMRG) simulations are presented for selected strongly correlated molecular systems addressing problems on Hilbert space dimensions up to $2.88\times10^{36}$.
\end{abstract}

\maketitle
%----------------------------------------------------------------------------------------

\section{Introduction}
\label{sec:intro}

In the past decades, numerical (classical) simulation has become an important part of both basic and applied research. This development has been made possible by enormous progress in High-Performance Computing (HPC) ~\cite{anzt2023high}, together with the development of numerical algorithms in simulating physical, chemical, biological, economical and ecological systems among many others.
In applying classical numerical methods to the interacting quantum systems, however, a fundamental limitation emerges: the so-called curse of dimensionality, that the computational effort scales exponentially with the dimension of the Hilbert space for systems described by multiparticle Schrödinger equations. Unfortunately, there is no known universal “fix” for this problem. 
~\cite{Schollwock-2005,Noack-2005,Verstraete-2008,Legeza-2008,Chan-2008,Schollwock-2011,Szalay-2015a,Orus-2014,Khoromskaiaab-2015,Baiardi-2020}.

As the design and mass manufacturing of efficient quantum computers are still subject of intense research~\cite{Ferrari_2021,PhysRevLett,quantum_sim}, the numerical simulations of quantum many body problems still rely on classical computation. 
It is the interplay of quantum and classical simulation and the delicate divide between them \cite{Xu-2023-herculean} that is the focus of massively parallelized tensor network state (TNS) algorithms designed for HPC infrastructures~\cite{Hager-2004,Stoudenmire-2013,Nemes-2014,Brabec-2021,Zhai-2021,Gray-2021,Lyakh-2022,Ganahl-2023}. 

Our TNS-based approach focuses on the development of massively parallel algorithms that are not only highly scalable and ideal to use in an HPC environment, but by building on the foundation of quantum many body physics and applied mathematics the number of required arithmetic calculations has been reduced by multiple magnitudes. As a result the exponential time cost of the simulations has collapsed into polynomial complexity.
 
In this work, we put an emphasis on one of the subclasses of tensor network state algorithms called density matrix renormalization group (DMRG) method~\cite{White-1992b}, but the presented algorithmic solutions are equally applicable for general TNS topologies. In such cases large-scale tensor operations can be substituted with multi-million vector and matrix operations, of which many can be executed independently. Through the exploitation of these (in)dependencies, arithmetic operations can be reordered and put into multiple tiers of groups corresponding to specific software and hardware layers ranging from low level CPU and GPU based SIMD execution to high level HPC scheduling. As for every tier we can execute all operations contained within the same group independently of all other arithmetics residing outside the group, mass scale parallelism can be individually achieved for each tier of groups. The resulting parallelization is the combination of each tier's own massive parallelization, thus with suitable hardware infrastructure exascale computing~\cite{Lyakh-2022} is becoming a reality for DMRG based quantum simulations.

The paper is organized as follows. In Sec.~\ref{sec:methods} we introduce methods developed according to various parallelization strategies and novel algorithmic solutions to achieve an efficient hybrid CPU-multiGPU kernel for simulations on HPC infrastructures.  
Sec.~\ref{sec:impl} is devoted to implementation details via the framework of the density matrix renormalization group method for general model Hamiltonians including long-range interactions.
In Sec.~\ref{sec:results} we present numerical benchmarks and scaling analysis for selected chemical systems
together with future possibilities and energy consummations.
Point-by-point conclusions, Sec.~\ref{sec:conclusion}, close our presentation. 

%------------------------------------------------

\section{Methods}
\label{sec:methods}

In pursuit of absolute performance, there is an evergrowing, relentless need for lighter, faster, more flexible, yet easier to deploy constructs of parallelization. In this section we introduce a few findings of our own; new methods and toolsets to aid us on our journey to better our software.

\subsection{Maze-Runner} 
In traditional producer-consumer models threads are casted into disjoint sets labeled as \textit{producers} and \textit{consumers}. The former's job is to break down the currently executing program's main task into smaller chunks of independantly solvable subtasks. Then, these task fragments are stored in some form of a buffer and are continuously fed to consumers, whose job is to solve said subtasks.

Ideally, producer and consumer threads can run in parallel, making this model --- despite its simplicity --- a highly effective path to create multithreaded workloads.

\subsubsection{Batched Type Task Processing} 
In theory the buffer of a perfect producer-consumer model is guaranteed to be not empty and not full at all times during use. If any of these two cases were to occur, either the producers or the consumers would be stalled. Thus, near-perfect thread allocation between producers and consumers are required.

Unfortunately the optimal distribution of threads is oftentimes problematic, especially when the relation between the difficulty of creating and solving tasks is volatile in nature. In such cases the streaming of tasks are not uniform, meaning task generation time can fluctuate rapidly and does not scale linearly with task size. Unless the threads are continuously re-distributed between producers and consumers, performance will suffer.  

Instead of implementing high-complexity dynamic scheduling systems relying on task specific optimizations, we present a more general approach that, instead of solving it, skips the problem entirely.

By creating a pool of general purpose threads, then ordering the pool to gather all currently available tasks before allowing the individual threads to consume any task, we can trade our scheduling problems with another set of problems. Mainly, the continuous flow of data is broken and we are expected to lose benefits related to pipeline based parallelizations.

Let us correct our approach by planting iterations into our model. Producing and consuming tasks in small batches reintroduces the dataflow, albeit with bigger, less frequent packages between pipeline elements. This also leads us to performance and utilization problems due to fragmentation of the pool of available tasks.

However when an implicit barrier\footnote{In parallel programming a \textit{barrier} is a construct that stalls certain threads, making them unable to go past a certain point in their execution --- that is, until the barrier is called upon from all interrelated threads.} is already present due to algorithmic reasons related to the subject\footnote{The base algorithm we want to parallelize.} of our parallelization, none of these poses a problem as the iterations of the parallelization model can be synchronized to the logic of the executing algorithm. In simple terms what this means is that for iterative algorithms this modification of the producer-consumer model can be implanted with effectively no overhead, since the framework enabling the algorithm's own iterations can be re-used as the main loop of the parallelization model. In a sort of way we are using the base algorithm's own inability to parallelize consecutive iterations to shadow the exact same bottleneck we introduced to our system by processing tasks in multiple batches.
\begin{figure}
 \includegraphics[width=0.48\textwidth]{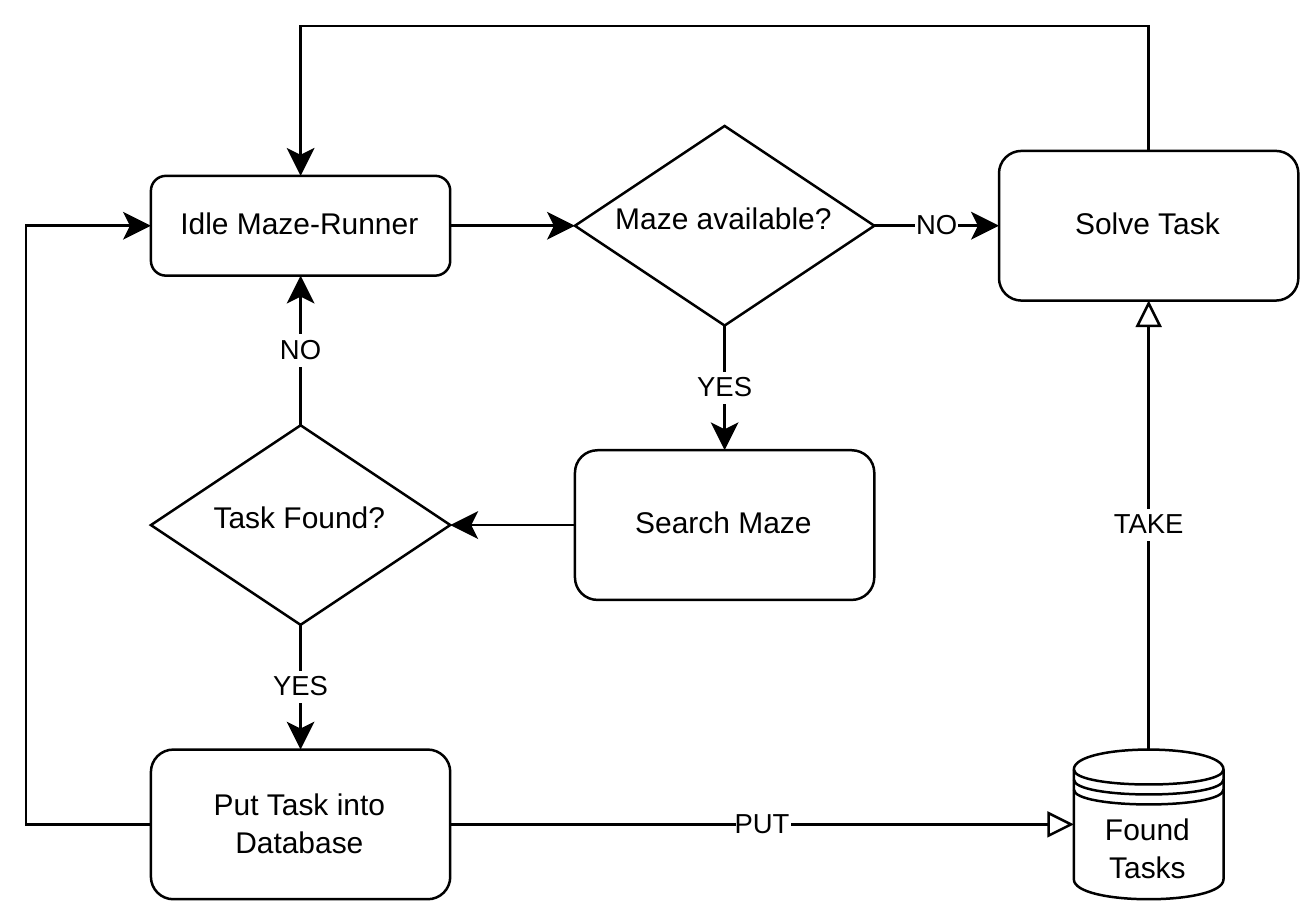}
   \caption{Life Cycle of a Maze-Runner Thread}
\label{fig:femoco}
\end{figure}

\subsubsection{Replacing Producers and Consumers with Maze-Runners} 
In cases where the generated tasks are products of interactions between complex systems, the process of producing tasks for the consumers starts to resemble a maze-like structure with all sorts of items hidden inside. When a thread enters, the outcome is questionable in a way that it is hard to predict what kind of task the thread will find and how long the search will take. Without advanced scheduling and preprocessing the threads are delving into the unknown. By introducing new terminology, let us simply call these uncharted territories as the \textit{maze}.

The threads will keep re-entering the maze until its misteries are all but revealed. While some threads are still inside, trying to find the last few tasks, the outsiders will no longer re-enter the maze. Instead, they will start solving the already gathered tasks. By doing so, a transition from production to consumption begins to take place. Since the exact same threads are used for both the production and consumption of tasks, intuitivaly we could call these producer-consumer threads \textit{maze-runners} instead.

As a generalization, maze-runners are all consumers of the higher tasks produced by the base algorithm itself. A consumer solves such higher task by generating yet another task, which is then re-introduced to the threadpool as a finding in the maze. What we get in the end is a recursive, self-feeding executor\footnote{\textit{Executors} are extended threadpools with scheduling and an interface for accepting tasks.}.

The performance benefit of our model comes from the fact that threads are not associated with the current recursion level, tasks are. On the contrary, threads can be fed with tasks from any level of recursion. This ensures a magnitude of thread utilization not feasable with classical producer-consumer based pipelines as these methods force threads into certain roles that restrict the range of acceptable tasks. While extended versions of these classic models can change the roles of threads when deemed necessary, introducing high complexity scheduling systems increases overhead as well as code complexity. Our approach removes the necessity of non-trivial scheduling, while at the same time allowing all threads unrestricted access to all available tasks inside the threadpool. 

\subsubsection{Notes on Implementation and Code Complexity}
Generally speaking, sequential loops with inner loops containing non-trivially gatherable subtasks are good candidates for the maze-runner variation of the classic producer-consumer model. 

In most cases the implementation is as simple as wrapping the parallelization model around the body of the algorithm's own iteration and providing information on what is considered to be a task, consequently creating an intermediate layer between the inside and outside of the inner loop. This way the parallelization model and the algorithm itself are only loosely connected, thus keeping such architectures relatively easy to implement and develop.

\subsubsection{Notes on non-CPU based Task Consumption}
The proposed model prioritises task-generation speed above all else, as all threads are initially allocated to task creation. When maze-runners are used to feed non-CPU processors such as GPUs or FPGAs, the risk of task-starving the devices is minimalized.

\subsection{Tree-Traversal Optimized Virtual Memory Addressing}
With the evergrowing computational power of hardware and scalable software, IO operations such as allocation, deallocation and copy are increasingly more difficult to implement\cite{Gelado2019} without bottlenecking the computational part of our algorithm. In this section we propose methods enabling us to create a system of virtual memory adressing, that not only enables us to allocate and deallocate in $\mathcal{O}(1)$ time, but can also dramatically reduce the number of copy operations needed due to its natural ability to cache and reuse intersections of consecutively used data groups.

\subsubsection{Data Dependency Trees}

\begin{figure}
\vspace{-\baselineskip}
\begin{algorithm}[H]
\begin{algorithmic}
\caption{TTcache - DFS on external database}\label{alg:TTcache}
\Procedure{load}{$data,cache,offset$}
\State $memcpy(cache+offset, data, size(data))$
\State $offset \gets offset + size(data)$ \Comment{visible from outside}
\EndProcedure
\vspace{0.5em}
\Procedure{unload}{$data,cache,offset$}
\State $offset \gets offset - size(data)$ \Comment{visible from outside}
\EndProcedure
\vspace{0.5em}
\Procedure{visit}{$v,database,cache,offset$}
\State $data \gets database.get(v)$
\State $load(data, cache, offset)$
\State $execute(v, cache)$ \Comment{solve tasks in $v$ locally}
\For{$child : v.children$}
    \State $visit(child, database, cache, offset)$
\EndFor
\State $unload(data, cache, offset)$
\EndProcedure
\vspace{0.5em}
\Procedure{TTcache}{$database, root$}
\State $cache \gets malloc()$ \Comment{allocate all available memory}
\State $offset \gets 0$
\State $visit(root, database, cache, offset)$
\EndProcedure
\end{algorithmic}
\end{algorithm}
\vspace{-\baselineskip}
\end{figure}

One of the naive solutions to memory management is to store all required data in memory at all times. When memory is abundant, it is not necessarily needed to improve upon this model, since we are already guaranteed to require only one copy per dataset, which is ideal from a performance perspective.

However, when the combined size of datasets exceed the size of allocatable memory, minimalizing the number of copy calls for each dataset can become excessively problematic. In simple cases using buffers and loading the data in chunks can solve the issue. Unfortunately, in complex cases --- in which no semantically equivalent restructuring of the algorithm exists, where the consecutive batches of data have a low enough overlap for the IO to be hideable behind the parallely running computation --- more sophisticated methods are required.

By reordering the computations related to particular datasets, it is often possible to place overlapping datasets next to each other during execution. This alone, however does not necessarily yield improved performance as the overlapping subsets of datasets can recursively overlap with other overlapping subsets of datasets. In other words, placing similar data next to similar data only works if similarity as a construct can be defined by a single attribute. In order to effectively cache overlapping sections of data, a toolset for managing multi-level dependencies between computation and data will be needed.

To represent the attributes of similarity by which datasets are linked to computation, we shall build a tree whose name we will choose as \textit{data dependency tree}. As every node contains an executable subprogram, by traversing the tree we execute the program itself. Contrary to structuring our program as a sequence of code blocks, this tree based view of our execution path enables us to store additional information, that otherwise would be cumbersome to deal with.

Specifically, by requiring child nodes to contain the same attributes as the parent, the entire subtree defined by the parent node can be executed without any required IO operations related to attributes within the parent node. This is evident, considering all data defined by the parent node's attributes are also present in all nodes of the subtree, thus for every subtree the root node's data overlaps all nodes within the subtree. By choosing frequently occuring attributes for higher levels of the graph, storing ancestor node data during subtree execution can lead to a sensible way of caching.

It is of great importance to understand, that generating a data dependency tree for a given problem is an ambiguous task. Especially because we have the freedom to cache data not required by the current computation. What this means is that a node might have ancestors whose dependencies are not fully realized by the node, but are included anyway as it enables the node to share common traits with an extended family\footnote{For example we might make non-electric engines dependant on electric engines and vice versa, even though it would not be necessary. As a benefit, we can process both electric and non-electric cars this way without constant IO operations related to engine type.}.

\begin{figure}
 \includegraphics[width=0.48\textwidth]{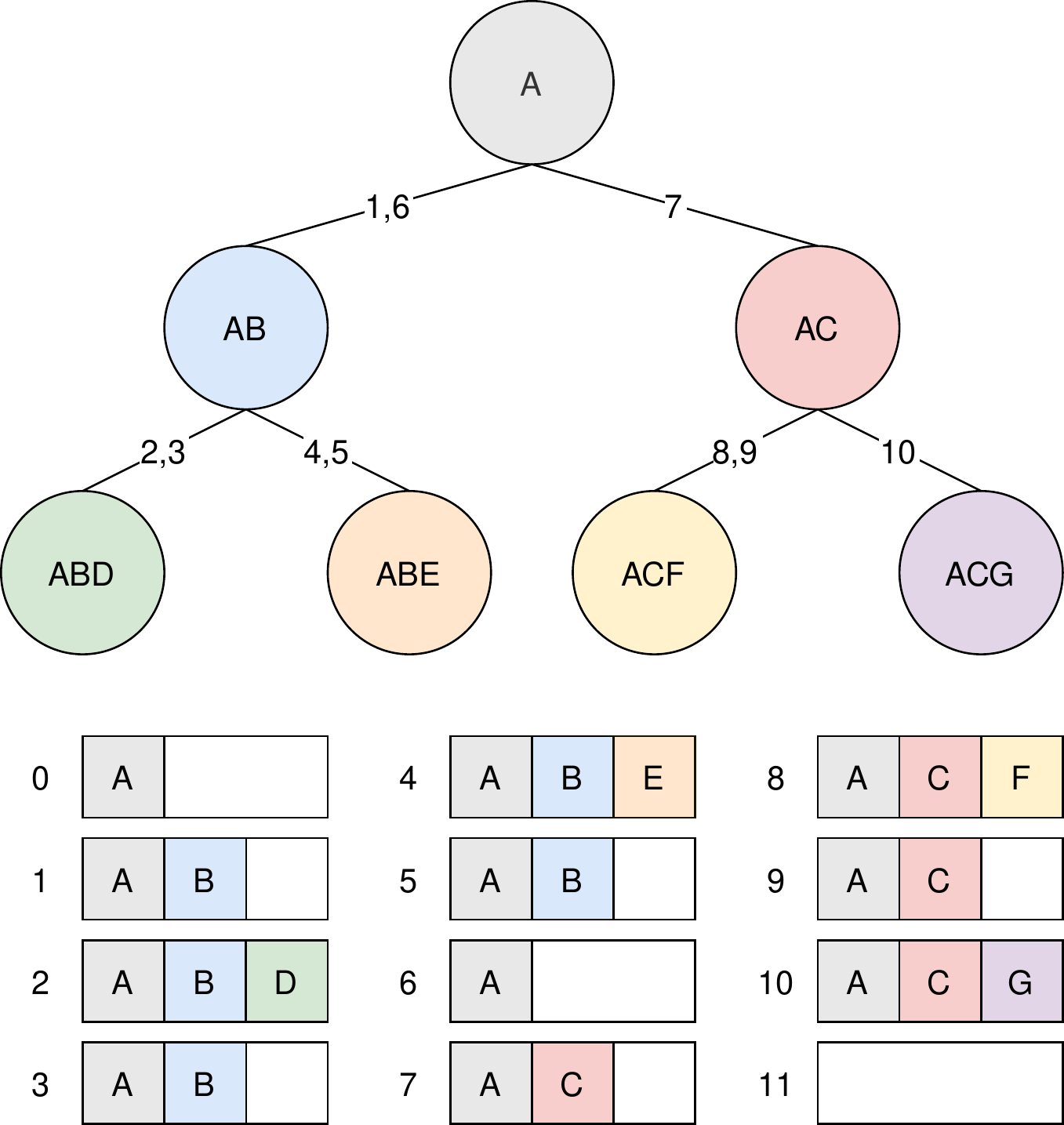}
   \caption{Buffering while Traversing the Data Dependency Tree. The numbers represent the order in which the vertices are visited. The arrays show the buffer's content for each step.}
  \label{fig:tree}
\end{figure}

\subsubsection{Fragmentation Free, Constant Time Pseudo-Allocations}
Data dependency trees can be pre-calculated. Also, in many cases it is possible to generate data with the tree like structure already present, thus rendering the creation of the dependency tree trivial. 

When used for caching, DFS\footnote{Depth-First Search is a graph traversal algorithm. In case of trees, DSF will always choose children over siblings.} traversal will chain ancestor nodes together, guaranteeing gap free, sequential writes to memory. When moving up the tree, the indexed part of the memory is simply shortened to fit only the current node's dependencies.

Sibling nodes are oblivious to each other's existence, and so will gladly overwrite previous buffer information starting at the end of the parent node's data. This behaviour is a welcome addition to our model as it renders both conventional allocations and deallocations unnecessary.

Data can be safely indexed as ancestor node data is always located at the front of the buffer and is not touched by the current node, since these are all current dependencies as well, and are used by the node. Overwriting the memory starting at the end of ancestral dependencies are irrelevant to all other nodes, because only descendants use data from previous nodes. Using DFS, we are guaranteed to have all these children already visited by the time this data is disregarded.

The described memory model provides gap-free, sequential write and read operations, moreover, no allocations and deallocations are required in the traditional sense. Allocation is as simple as returning the beginning of the currently unindexed portion of memory. Deallocations and memory leaks do not exist in this system, because indexing is node dependent and is disregarded by all non-descendent nodes.

We call the memory model described above \textit{Tree-Traversal Optimized Virtual Memory Addressing} --- or \textit{TTcache} for short. Fig.~\ref{fig:tree} shows a simple example of TTcaching. The outline of our procedure is available in the form of Alg.~\ref{alg:TTcache}. As the basis for our program, we traverse the data dependency tree using DFS. Since the external data required for the whole operation is too enormous and cannot be fitted into local memory, some form of buffering is required. However, by loading the data in multiple chunks we run the risk of calling overlapping IO operations. Using TTcaching, such redundancies are eliminated, thus easing the burden of data transfers between external and internal memory spaces.
\begin{figure}
  \includegraphics[width=0.48\textwidth]{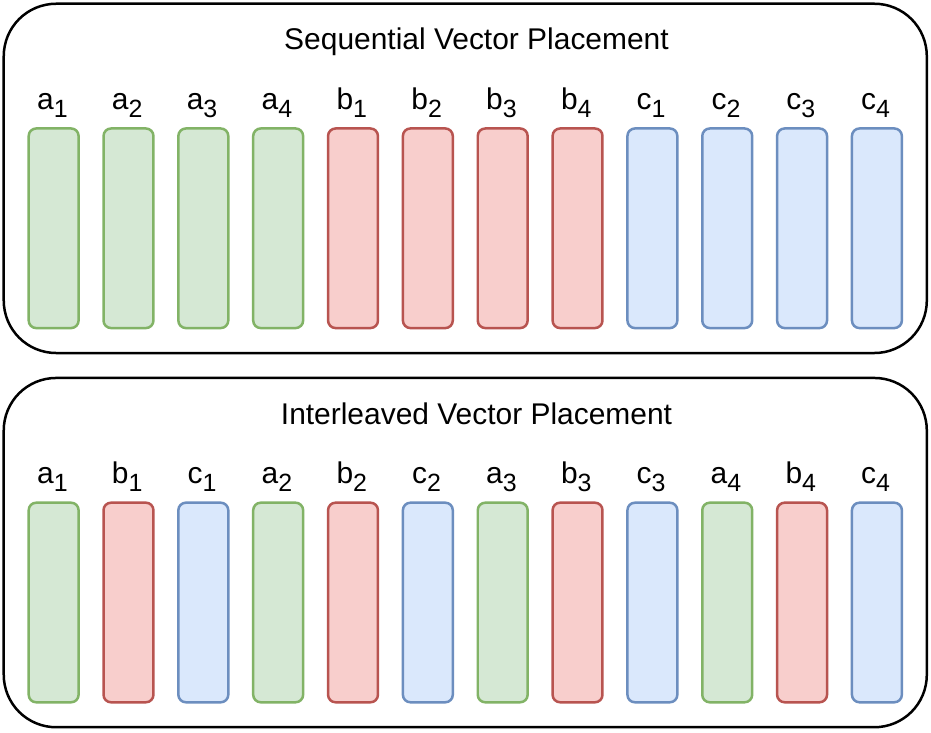}
  \caption{Vectorized output of a Strided Batched GEMM operation. Normally, output vectors belonging to the same matrix are in a sequential order (top). However, interleaving the vectors of different matrices (bottom) is possible by altering the leading dimensions and stride values of the output matrices.}
  \label{fig:sbmm4s}
\end{figure}

\subsection{Strided Batched Matrix Multiplication for Summation}
SIMD workloads have a tendency\cite{DONGARRA2017495} to perform poorly when bombarded with a high amount of small jobs. This is understandable, considering data-level parallelism cannot reduce the number of instructions a thread has to crunch through. Also, smaller tasks do not offer much to distribute among threads. To improve performance on hardware that is incapable of instruction-level parallelism, aggregating smaller jobs into bigger ones appears to be a step in the right direction. With fewer, bigger SIMD parallelized blocks we not only reduce the number of sequential instructions, but also make it easier for schedulers to efficiently utilize threads, as more data without more instructions means more parallelism. 

For aggregation of matrix multiplications, both Intel and NVIDIA has implemented solutions. Batched GEMM\footnote{GEneric Matrix Multiplication, often available in single-precision, double-precision, real and complex flavours.} functions work in a similar fashion as their non-batched counterparts. Instead of two input and one output matrices, they operate with two arrays of input matrices and one array of output matrices. A strided variant is also often included, which does not require arrays. Instead, these functions use offsets to determine the memory addresses within a batch. Strided Batched variants promise to eliminate the overhead caused by arrays.

Unfortunately, because each multiplication is calculated separately, the results are also presented as separate matrices. This can pose a problem when the goal is to calculate the sum of multiple matrix multiplications. While it is fairly easy to solve the reduction\footnote{In vectorized workloads the reduction problem arises when the algorithm outputs multiple results due to it's SIMD nature, and not because more than one result is required. In these cases the result vector has to be reduced to a single entity. Sum reduction is a reduction problem, where the reduction happens by applying summation.} problem, these are unnecessary operations that would not be present if we were to sequentially apply the matrix multiplication while accumulating the results in an output matrix that is shared among GEMM calls.

The first step in solving the problem --- and enabling batched typed matrix multiplication without the overhead of sum reduction --- is to understand the way matrices are stored. In column-major ordering matrices are arrays of vectors representing matrix columns, while in row-major ordering these vectors are interpreted as the rows of the matrix. What matters to us is that in both cases matrices are arrays of vectors. 

The second pillar of our method is the relation between sequentially stored matrices. Adjacent matrices, depending on their ordering, can be interpreted as an either horizontally or vertically concatenated larger matrix. The reason we cannot operate on two of these aggregated matrices is that both batches of matrices are concatenated the same way. For valid multiplication the left batch has to be horizontally concatenated, while the right batch requires vertical concatenation. This way the extra length and height of these large matrices will dissolve during multiplication, resulting in a normal sized matrix. Since we used the intrinsic sum reduction of the matrix multiplication itself, no further calculations are needed.

Finally, the only tool missing is the way to flip the concatenation of a matrix batch. This is where we reach for strided batched matrix multiplications again. While these operations cannot produce a single result, we can however create intermediate batches of matrices with flipped concatenations. In order to do this all that has to be done is setting the output matrix offsets in a way that the vectors of output matrices will be interleaved. As it is visible in Fig.~\ref{fig:sbmm4s}, vectors from different matrices, but with the same index, will take place next to each other. These vectors will form the long vectors of the aggregated matrix during sequential memory reads.

By consecutively applying strided batched matrix multiplication and traditional matrix multiplication on aggregated large matrices, we can perform batched type chained matrix multiplications without sum reduction at the end. We call this method \textit{Strided Batched Matrix Multiplication for Summation} --- or \textit{SBMM4S} for short.

%------------------------------------------------

\section{Implementation details}
\label{sec:impl}

Our algorithmic developments discussed in the previous sections will be presented for the density matrix renormalization group (DMRG) method~\cite{White-1992b} that is a special variant of the tensor network state (TNS) algorithms.~\cite{Schollwock-2011,Noack-2005,Chan-2008,Szalay-2015a,Orus-2014,Baiardi-2020}
We focus on a very general form of the Hamiltonian operator, implemented in our code ~\cite{dmrg-budapest},  
that can treat any form of non-local interactions related to
two-particle scattering processes. The corresponding
Hamiltonian can be written in the form
\begin{equation}
\mathcal{H} = \sum_{ij\alpha\beta} T_{ij}^{\alpha\beta} 
                c^\dagger_{i\alpha}c_{j\beta} +
                \sum_{ijkl\alpha\beta\gamma\delta} 
      {V_{ijkl}^{\alpha\beta\gamma\delta}
      c^\dagger_{i\alpha}c^\dagger_{j\beta}c_{k\gamma}c_{l\delta}},
\label{eq:ham}
\end{equation}
where the indices
$\alpha,\beta,\gamma,\delta$ label internal degrees of freedom, like spin or isospin.
The
operators $c^\dagger_{i\alpha}$ or $c_{i\alpha}$ usually denote spin ladder or
fermion creation and annihilation operators.
Indices $i,j,k,l$ label modes, which can
be, for example, lattice sites in real-space
representation~\cite{White-1992b}, band and
momentum indices in momentum space representation~\cite{Xiang-1996},
molecular orbitals in quantum chemical applications~\cite{White-1999},
spinors in relativistic problems~\cite{Knecht-2014},
proton and neutron orbitals in nuclear structure theory~\cite{Dukelsky-2004,Legeza-2015},
modes of particles confined in a Harmonic traps
~\cite{Legeza-2018a,Shapir-2019},
or Kohn--Sham orbitals 
~\cite{Barcza-2020} among many others.
In order to boost the performance of the method, various algorithmic
solutions based on concepts of quantum information theory have been utilized~\cite{Legeza-2003a,Legeza-2003b,Barcza-2011,Fertitta-2014,Krumnow-2016}.

In this work, instead of providing benchmark results for the above mentioned systems we present a detailed analysis of various scaling properties for two chemical compounds.
In addition, our massively parallel, multi-GPU accelerated architecture is utilized via the diagonalization of the so-called effective quantum many body Hamiltonian and the so-called renormalization procedure. These two usually correspond to 95\% of the total execution time. The details of the algorithm can be found in various review articles \cite{Schollwock-2005,Schollwock-2011,Noack-2005,Szalay-2015a,Chan-2008,Orus-2014,Baiardi-2020}.

From the perspective of computer science, the key aspect of TNS/DMRG
algorithms is the fact, that the exponential scaling governing exact diagonalization can be
reduced to a polynomial form, and the underlying tensor and matrix algebra, known as tensor product factorization, can be organized into several million of independent operations (tasks). 
Therefore, the dense matrix operations are performed in parallel 
according to the so-called quantum number decomposed representations (sectors). 
The size of the full matrices, denoted as DMRG bond dimension, $D$, determines the accuracy of the calculations and at the same time the required
computational complexity. The overall scaling of the DMRG is $D^3N^4$ where $N$ stands for the number of modes, i.e., for the system size. The memory requirement is proportional to $D^2N^2$.

The diagonalization of the effective Hamiltonian is performed iteratively via the L\'anczos or Davidson algorithms, usually accounting for 85\% of the total execution time.
This involves a series of matrix multiplications and summations. 
The runner up for longest running procedure is the transformation of the operators, also know as renormalization or blocking,  which is decomposed into a series of tensor product operations, matrix multiplications and summations. The renormalization step is responsible for ~10\% of the total execution time.

\subsection{DMRG Integration}
\label{ssec:dmrg-integration}

Here we present basic implementation details of the various algorithmic solution discussed in Sec.~\ref{sec:methods}. We follow the traditional DMRG picture instead of the MPS based description, emphasizing the fact that they are equivalent.
In addition, we summarize only those technical aspects which are relevant to parallelization issues. 

\subsubsection{Quantum number dependent tensor library}
In the DMRG algorithm the modes of a network are partitioned into subsystems (blocks), and the algebra is performed based on tensor products of the operators represented on these subsystems~\cite{Schollwock-2005,Noack-2005,Schollwock-2011,Orus-2014,Szalay-2015a}.
In addition, the matrix and tensor representations of the operators are decomposed to sectors based on quantum numbers, i.e., a full matrix is stored according to row-column quantum number sector pairs and the corresponding dense matrix ~\cite{Nemes-2014,Szalay-2015a,Brabec-2021}. This is similar to the sparse representation of matrices, where non-zero scalar elements are stored according to the corresponding row-column index pairs. All operations of the underlying DMRG linear algebra is developed via such sector representation. In our tensor library ~\cite{tenlibol}, the sector dependent dense matrices of the same operator type  
have a third index as well, labeling their positions in the given subsystem. All tensor operations are implemented according to a four-level hierarchy: 
\begin{enumerate}    
\item
Based on the sector description tables
  (list of sector row-column pairs) only the possible 
  output sector pairs are determined (sector-table).
\item
Based on the sector description tables the possible output sector pairs are determined together with a table storing all sector combinations of the input operators and the corresponding output sectors of the resulting operation
(task-table).
\item
Besides steps one and two the actual operation is also performed. 
\item
(Optional) Besides steps one to three the same operation is formed in full form representation and a self-check is performed.
\end{enumerate}

\subsubsection{Independent operational tasks}

For long range interaction, operators appearing in the Hamiltonian Eq.~(\ref{eq:ham}) are distributed among the various subsystems of the network, also known as operator factorization or matrix product operator (MPO) representation~\cite{Schollwock-2011}. In addition, partial summations or pre-contractions are performed to reduce the 
overall complexity of the MPO representation of the Hamiltonian 
from $N^4$ to $N^2$ (for details see for example ~Ref.~\cite{Xiang-1996,Szalay-2015a}). 

In the two-site DMRG topology the modes are partitioned into four subsystems, where the so called left and right blocks, collection of several modes, are mediated by two intermediate subsystems representing single modes. \cite{White-1992a}.
Therefore, the matrix vector multiplication in the iterative diagonalization via the Davidson or
L\'anczos algorithm is obtained from an accumulated sum of  
four matrix multiplications along four distinct dimensions of a four-dimensional tensor.
For models when the local Hilbert spaces of the two-intermediate modes are decomposed into one dimensional sectors based on quantum numbers -- like in the model systems studied in this paper --, i.e., when only scalars are stored according to row-column sector index pairs for operators of the two-intermediate modes,
the contributes of these can be precalculated leading to an overall scalar multiplication of each operation elements stored in the task-table. This reduces the problem to a series of matrix multiplications in the form 
\begin{equation}
B \coloneqq B + \alpha \sum_{i=1}^p L_i A R_i^T\,,
\end{equation}
where $p$ stands for the number of independent tasks,
$\alpha$ is a precalculated constant, $L$ and $R$ label left and right block operators, and
$A$ and $B$ are the matrix representations of the quantum many body wavefunction in quantum number sector decomposed form. 
For more details see Refs.~\cite{Schollwock-2011,Szalay-2015b} and the pseudo-code Alg.~\ref{alg:SBMM4S} discussed in the next section.

In our implementation, each of such operator combination is stored in a table (operator-table) and each operator combination supplied with the corresponding task-table. Therefore, there are three main loops that must be executed: loop over the rows of the operator-table, for each row a loop over the rows of the corresponding task-table and finally a loop over the position index within the subsystem blocks. The product of these provides the number of independent operations that must altogether be executed to perform a given algebra~\cite{Nemes-2014,Brabec-2021}. Relying on the four-level hierarchy discussed above before a given sequence of operations is executed, first the level-one and level-two procedures are performed in order to determine the corresponding tables. This means that execution of the independent tasks given by the tables follows only after such initialization procedure is completed, which makes our implementation ideal for HPC infrastructures
via dynamic scheduling protocols.

\subsubsection{Constructing the Data Dependency Tree}
As outlined in Sec.~\ref{sec:methods}, there are multiple methods for organizing the underlying DMRG algebra into independent tasks, with each varying in asymptotic space, IO time and compute time complexities. Therefore, we have developed various algorithmic solutions to generate the data dependency tree presented in Sec.~\ref{sec:methods} depending on what parallelization strategy is used for a given problem \cite{Nemes-2014,Brabec-2021,Ganahl-2023}.

Since GPU devices are handled at a lower abstraction level, all mathematical aspects of the DMRG algorithm is considered on the host (CPU) side. GPU devices are used only for executing basic algebraic operations in large batches. These highly vectorized operations are defined by the so called task tables, which themselves are the direct output of host side preprocessing. Although these initial steps implement complex mathematical models, they are computationally lightweight as no actual data transformation occurs. Such architecture enables us to use MATLAB, a high level interpreter based language for modeling complex systems, while number crunching remains close-to-the-metal thanks to the entire computational part being implemented in native C++. Low overhead traversal between the two worlds are made possible by MATLAB's MEX interface.

DMRG calculations for model Hamiltionians with long range interactions require large amount of data stored in RAM that far exceeds current GPU memory sizes. We have found that the parallelization scheme where the individual tasks are formed according to the operator factorization of the Hamiltonian (rows of the operator-table) is more efficient, than the model in which all tasks are grouped by the quantum number output sectors of the wave function\cite{Nemes-2014,Brabec-2021}. This allowed us to reduce the required IO operations tremendously by applying novel procedures presented in Sec.~\ref{sec:methods}.

During computation, a thread is assigned for each GPU device with the main goal of pushing both compute and IO operations into the associated device's CUDA streams. Results are collected by the same threads, making D2H\footnote{Device-to-Host: copying memory from VRAM to RAM} data retrieval also parallel, thus enabling a level of RAM and PCI-E lane saturation not feasible with a single card. Ultimately, collected results are merged into the quantum number sector components of the wave function at the host side. Fine grain mutual exclusion is used to ensure thread safety for simultaneous write operations. Since the corresponding memory segment for each lock is significantly smaller than the span of a given sector, high throughput with minimal clashing between threads is possible.

As the number of threads dedicated to GPU devices increase, the host is left with fewer threads for local computations. Considering the heavy lifting is now handled by accelerators, the host is left with nothing, but leftovers to compute. These tiny operations are too small to make their offloading to other devices worthwhile. Luckily for us, it also means inplace execution is almost instantaneous and, as such, our CPU-GPU hybrid model is fairly insensitive to low CPU core count.

\subsection{Competing Algorithms}
There are two schools of thought in parallel programming\cite{2schools}. Considering programs consist of data and instructions, it is only natural to think about data-level and instruction-level parallel models. While both ideas are broad terms consisting of multiple approaches on their own, it is oftentimes best to marry these concepts\cite{2choolstask} in a way that the currently used hardware can be pushed to its limits when encountering a particular type of workload. 

In the following sections we will compare four algorithms. Two reference designs are built using industry standard high performance linear algebra library \textit{Intel MKL}\cite{mkl}. On top of these two algorithms, we implemented our own methods, supplementing or replacing Intel MKL counterparts as neccessary.

To ensure fairness of testing, all four algorithms stem from the same base model, meaning the code for the compute algorithm is exactly same. All versions contain the same level of optimization which includes optimal parenthesization of matrix-chain multiplications\cite{chikalov2011sequential}, buffer re-usage, dynamically choosing between inplace and buffer-based calculations, replacing IO with pointer arithmetics whenever possible, and many more.

\subsubsection{Multithreaded SIMD Matrix Operations}
For our first reference design we use no explicit parallelization construct in our code. Instead, we link our software with library \textit{Intel Threading}, which contains function definitions for the Intel MKL API with inherent parallelization. Every time a vector or matrix operation is called, the API will refer to a built-in, pre-compiled function with internal OpenMP multithreading.

By executing vector and matrix operations one-by-one and distributing algebraic structures to different threads intead, we effectively get SIMD\footnote{Single-Instruction Multiple-Data, vectorized tasks usually fall into this category} typed parallelization. Different threads are executing the same instruction, but on different parts of the data.

\subsubsection{Task Parallelization with Dynamic Scheduling}
A different approach is to leave vector and matrix operations in one piece, and instead feed different operations to each thread. By doing so, we create a more abstract, instruction-level implementation. Unfortunately, with the increase of abstraction we have now reached the peak of Flynn's Taxonomy\cite{shinde2015architectures}, meaning this version of our program will be strictly MIMD\footnote{Multiple-Instruction Multiple-Data, not to be confused with instruction-level parallelization, as the later can also mean MISD, which stands for Multiple-Instruction Single-Data} and cannot be transported to SIMD hardware such as GPUs.

We use an OpenMP parallel pragma on the outer loop of the maze, essentially assigning a unique section of the maze to each thread. There exist no separate task creation and consumption here. Once a thread enters the maze, it will not rest until the given section is cleared. This model bears a resemblance to thread-pool based parallel for-cycles and traditional executors.

In order to make our implementation more flexible and more fit for maze solving, we use Intel's implementation of OpenMP dynamic scheduling\cite{duran2008evaluation}. It enables the thread-pool to compensate for differences in task difficulty and thread execution speed by streaming tasks to threads on the fly instead of static distribution. 

\subsubsection{Implementing Maze-Runner using Intel OpenMP}

The Maze-Runner version extends the functionality of the previous MIMD version by transforming the parallel loop into task creation and allowing task consumption at a later time. This results in tuneable task granularity and also enables out of order\footnote{\textit{Out of order execution} usually refers to hardware level dynamic scheduling of instructions, however in this context by \textit{out of order} we plainly mean the order of task consumption does not have to match the order of task gathering.} task scheduling.

By using Maze-Runner threads, task creation is akin to classical producer-consumer models, but without the need for multi-role threading and the scheduling complications that arise from the higher complexity of MPMD\footnote{Multiple-Program Multiple-Data, traditional producer consumer models and pipelines run different subprograms on different threads. Such architectures are classified as MPMD. Meanwhile, our Maze-Runner model enjoys the simplicity of Single-Program Multiple-Data architectures, while still providing customizable tasking.} models.

\subsubsection{Implementing SBMM4S and TTCache using CUDA}

\begin{figure}[t]
\vspace{-\baselineskip}
\begin{algorithm}[H]
\caption{SBMM4S with tensor operators (CUDA)}\label{alg:SBMM4S}
\begin{algorithmic}
\Require 
    \begin{tabular}{ l l }
     $\alpha \in \mathbb{R}$ \\ 
     $A \in \mathbb{R}^{m \times n},$ & $B \in \mathbb{R}^{q \times r}$ \\  
     $L \in \mathbb{R}^{q \times m \times p},$ & $R \in \mathbb{R}^{r \times n \times p} $
    \end{tabular}
\Ensure $B \coloneqq B+\alpha(\sum_{i=1}^{p}L_i*A*R_i^\top)$
\State \hrulefill \vspace{0.5em}
\State \textcolor{darkgray}{// Step 1: batched GEMM with interleaved output}
\State
    $cublasDgemmStridedBatched($ \\ 
    \begin{tabular}{ l l }
        \hspace{2em} $handle,$        & \textcolor{darkgray}{// handle}\\
        \hspace{2em} $CUBLAS\_OP\_N,$ & \textcolor{darkgray}{// transa}\\
        \hspace{2em} $CUBLAS\_OP\_T,$ & \textcolor{darkgray}{// transb}\\
        \hspace{2em} $m, r, n,$       & \textcolor{darkgray}{// m, n, k}\\
        \hspace{2em} $1.0$            & \textcolor{darkgray}{// alpha}\\
        \hspace{2em} $A, m, 0,$       & \textcolor{darkgray}{// A, lda, strideA}\\
        \hspace{2em} $R, r, r*n,$     & \textcolor{darkgray}{// B, ldb, strideB}\\
        \hspace{2em} $0.0$            & \textcolor{darkgray}{// beta}\\
        \hspace{2em} $temp, m*p, m,$  & \textcolor{darkgray}{// C, ldc, strideC}\\
        \hspace{2em} $pos$            & \textcolor{darkgray}{// batchCount}\\
    $);$
    \end{tabular}
\vspace{0.5em}
\State \textcolor{darkgray}{// Step 2: GEMM for concatenated vectors}
\State
    \begin{tabular}{ l l }
    $cublasDgemm($ \\ 
        \hspace{1em} $handle,$        &  \textcolor{darkgray}{// handle}\\
        \hspace{1em} $CUBLAS\_OP\_N,$ &  \textcolor{darkgray}{// transa}\\
        \hspace{1em} $CUBLAS\_OP\_N,$ &  \textcolor{darkgray}{// transb}\\
        \hspace{1em} $q, r, m*p,$     &  \textcolor{darkgray}{// m, n, k}\\
        \hspace{1em} $\alpha$         &  \textcolor{darkgray}{// alpha}\\
        \hspace{1em} $L, q,$          &  \textcolor{darkgray}{// A, lda}\\
        \hspace{1em} $temp, m*p$      &  \textcolor{darkgray}{// B, ldb}\\
        \hspace{1em} $1.0$            &  \textcolor{darkgray}{// beta}\\
        \hspace{1em} $B, q$           &  \textcolor{darkgray}{// C, ldc}\\
    $);$
    \end{tabular}
\vspace{0.5em}
\State \hrulefill \\ \textit{For detailed descriptions of the abbreviations used above, we guide the readers to the NVIDIA cuBLAS documentation.}
\end{algorithmic}
\end{algorithm}
\vspace{-\baselineskip}
\end{figure}

For our final algorithm we present a combination of both data-level and instruction-level parallelization by extending our previous implementations yet again by introducing full support for multiGPU systems.

Furthermore, the Maze-Runner CPU threads themselves are now capable of mixed MIMD and SIMD execution by allowing nested multi-level OpenMP threading. This includes explicit pool creation, handling and tasking as well as implicit second level constructs of parallelization in the form of threaded vector algebra.

This new hybrid CPU threading is merged with manually handled C++11 threading dedicated to asynchronously feeding an array of CUDA streams belonging to multiple GPU cards. Device side buffering is handled by our in-house TTcaching method showcased in Alg.~\ref{alg:TTcache}.

Computations are SBMM4S accelerated, while at the same time maintaining full compatibility with the CUDA library. As shown in Alg.~\ref{alg:SBMM4S}, this is all possible without the need to resort to custom written kernels, since the result vectors can be generated in a suitable order by manipulating certain offset values of cuBLAS kernels. In addition, our implementation also features partial SBMM4S support and traditional execution pathways as fallback, in case requirements are not all met.

In the end, the resulting algorithm is a multi-paradigm superhybrid featuring massive parallelization at multiple layers of software and hardware.

%------------------------------------------------

\section{Results}
\label{sec:results}

\subsection{Model systems}

In this section, we present benchmark results for two selected quantum chemical model systems. First results will be shown for the F$_2$ molecule in a CAS(18,18) orbital space~\cite{Schafer1992}
which problem corresponds to the current limit of exact diagonalization 
to obtain the full-CI energy
on HPC supercomputers, i.e., it describes the correlation of 18 electrons on 18 orbitals leading to a Hilbert space with dimension $9.075\times10^9$.
Next, results for the FeMoco will be shown which metal centered multi-reference problem is in the focus of modern quantum chemistry 
\cite{Reiher-2017,Li-2019,Kai-2020,Brabec-2021,Friesecke-2022b}
due to is important role in nitrogen fixation
\cite{Hoffman-2014}. 
Here much larger model spaces, i.e., CAS(54/54)
and CAS(113/76) introduced in Refs.~\cite{Reiher-2017}
and \cite{Li-2019}, respectively,
will be considered  with a full Hilbert space dimension of $2.485\times10^{31}$ and
$2.88\times10^{36}$.
These latter problems also server good benchmark systems regarding
parallelization issues as various reference data obtained by different methods are also available in the literature \cite{Kai-2020,Brabec-2021} utilizing, for example, up to $10^4$ CPU cores.
For these systems  accurate 
electronic structure calculations require large scale DMRG simulations,
with very large bond dimension $D$,
where massive parallelization is mandatory in order to bring computational time to a reasonable range \cite{Nemes-2014,Brabec-2021,Zhai-2021}.

Below we present our benchmark results 
as a function of the DMRG bond dimension $D$ and the number of GPU devices. For the F$_2$ molecule we used three DMRG sweeps leading to a relative accuracy of 10$^{-10}$ in the ground state energy while for FeMoco CAS(54,54) seven sweeps have been utilized to get converged energy in the error margins of $10^{-3}$ according
to available reference data ~\cite{Li-2019,Kai-2020,Friesecke-2022b}. Here we recall, that a DMRG sweep includes $N$ individual DMRG iteration cycles each decomposed to a series of algorithmic subprocedures executed sequentially. We also emphasize that we provide data for the total execution time measured via the sweeping procedure, while in various previous works measurements have been performed usually for the best DMRG configuration, i.e., when the size of the left and right block is half of the system size. 

Since the computational time of a full DMRG iteration cycle is determined mainly by the diagonalization of the effective Hamiltonians and by the renormalization steps, ~\cite{Nemes-2014,Brabec-2021} we will
focus on these two algorithmic procedures. 
In general, calculation of expectation values of operators and other measurable quantities based on the matrix product state (MPS) wave function exported from the DMRG are performed as post-DMRG procedures.

Here we remark that in many HPC infrastructures the computation time is limited to shorter time periods (often a one-day limit is enforced), thus the algorithm must include checkpoints from where it can be restarted in case if the total computation time exceeds this limit. This is achieved by saving the MPS data to HDD storage space which can be reloaded when the DMRG calculation is restrated from the last iteration step.   

\subsection{Multithreaded Diagonalization}

In this section we present results for the diagonalization of the effective Hamiltonian with paralellization solely relying on CPU core pinned\footnote{Pinning threads to CPU cores refer to a 1-on-1 relation between threads (software) and cores (hardware). Thread count cannot exceed core count and, ideally, the scheduler does not move threads between cores, hence the name "pinned".} Maze-Runner threads discussed in Secs.~\ref{sec:methods} and ~\ref{sec:impl}. For now, our focus is on single node calculations, therefore, the combination of our currently presented CPU and CPU-GPU hybrid solution with our Message Passing Interface (MPI) based implementation will be showcased in subsequent work.

\begin{figure}
  \centering
  \includegraphics[width=0.48\textwidth]{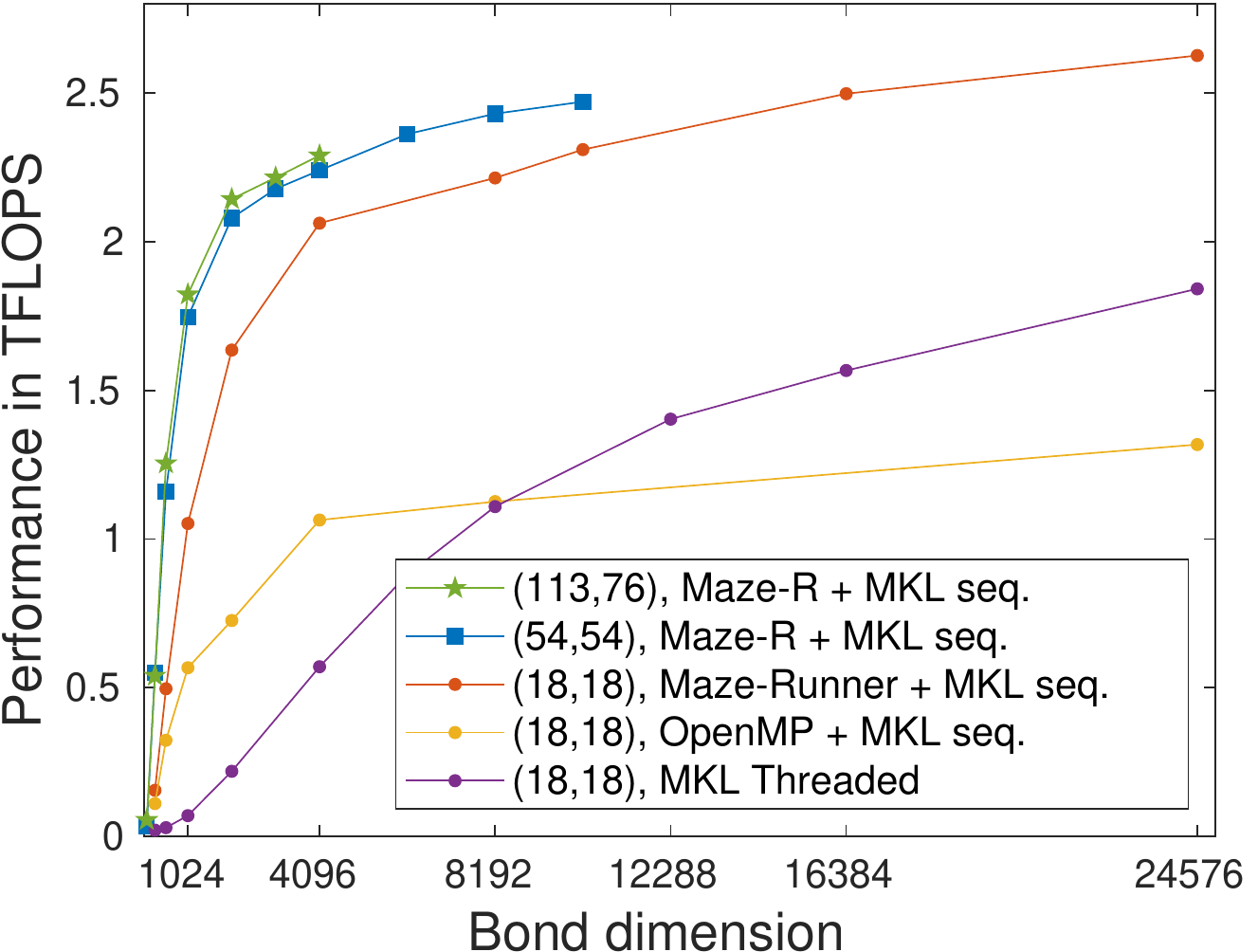}
  \caption{Performance measured in TFLOPS for the F$_2$ and FeMoco chemical systems for CAS(18,18) and CAS(54,54) orbitals spaces, respectively,
  as a function of the DMRG bond dimension on a dual Intel(R) Xeon(R) Gold 5318Y CPU system with $2\times 24$ physical cores running at 2.10 Ghz.}
 \label{fig:perf_cpu}
\end{figure}
Fig.~\ref{fig:perf_cpu} demonstrates the performance of various CPU based parallelization schemes derived from the diagonalization of the effective quantum many body Hamiltonian as a function of the DMRG bond dimension values.
Measurements are performed on a dual Intel(R) Xeon(R) Gold 5318Y CPU system with 2x24 physical cores running at 2.10 GHz.

By linking an otherwise sequential algorithm against Intel's threaded MKL libraries, the resulting program gains the ability to harness multiple cores simultaneously. This is possible as the library contains strictly high level matrix operations as its callable subroutines, with constructs of parallelism within the function definitions.

As expected, executing individual matrix operations sequentially does not cope well with small matrix multiplications. With not much to chew on, or should we say, to parallelize, the speedup from individually multithreaded matrix operations is not enough to significantly increase throughput and, alas, performance suffers the burden of millions of sequentially running small-scale operations. While the algorithm seems to continuously benefit from ever growing matrix dimensions --- enough to even take on one of its MIMD cousin --- for extremely large $D$ values the MKL based parallelization demonstrates an awfully slow climb towards the ballpark estimate of 2 TFLOPS.

Alternatively, single threaded MKL subroutines may be brought into play in conjunction with OpenMP directives, forming parallel loops in which MKL calls reside. A much improved performance is observed for both smaller and intermediate $D$ values. It, however, saturates at a significantly lower threshold value, due to the loop body performing sub-optimally as an independently executable task.

In contrast, Maze-Runner threads do not suffer from such illness. With full support for custom tasking akin to producer-consumer models, task creation is decoupled from the control-structures\footnote{Control-structures are the basic building blocks of programs. Common control-structures include: \textit{sequence}, \textit{loop} and \textit{conditional} } of the underlying mathematical model. This enables for finer load balancing and, ultimately, much faster saturation, especially for larger CAS spaces. In addition, the algorithm quickly reaches a performance level that seems unattainable by its competitors, even at much higher bond dimensions. 

\subsection{MultiGPU Accelerated Diagonalization}

\begin{figure}
  \centering
  \includegraphics[width=0.48\textwidth]{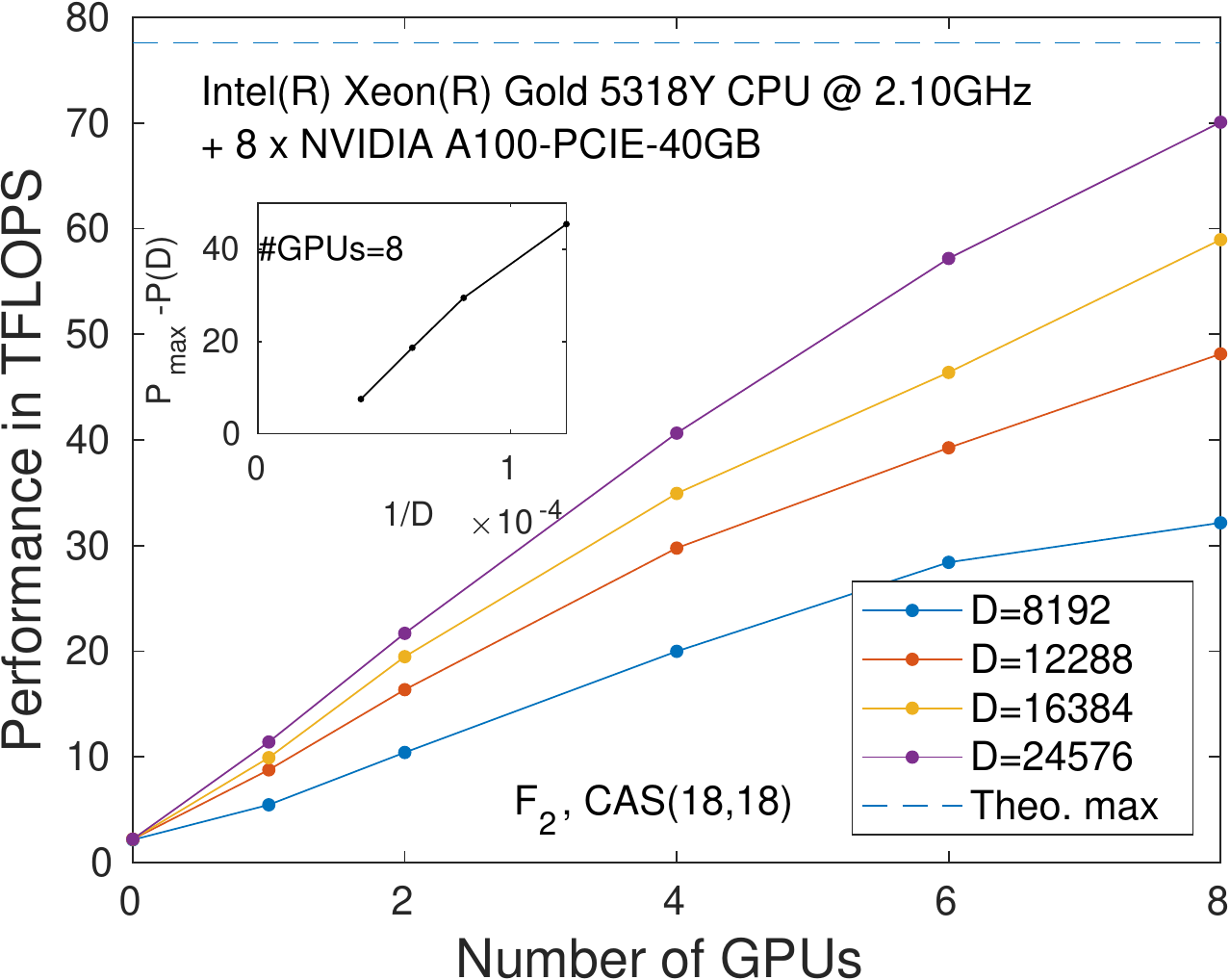}
  \caption{Performance measured in TFLOPS for the F$_2$ molecule, corresponding to CAS(18,18) as a function of the number of GPU devices for various fixed DMRG bond dimension values. Calculations have been performed on a dual
  Intel(R) Xeon(R) Gold 5318Y CPU system with 2x24 physical cores running at 2.1 GHz compiled with eight NVIDIA A100-PCIE-40GB GPU units.
  The inset shows the scaling of the performance with respect to the estimated theoretical maximum, $P_{\rm max}$ as an inverse of the DMRG bond dimension for eight GPU devices.
  }
 \label{fig:perf-gpu}
\end{figure}

Benchmark results are presented in Fig.~\ref{fig:perf-gpu} for our CPU-multiGPU hybrid model, which we discussed in Secs.~\ref{sec:methods} and ~\ref{sec:impl}. Performance measurement are derived from the diagonalization of the effective quantum many body Hamiltonian as a function of the number of GPU devices for various $D$ values.

In comparison to the CPU-only procedure, using even just a single GPU device can lead to a factor of two to four speedup in the performance.
With multiGPU configurations, the performance scaling in relation to the number of GPUs shows a closely linear trend. The slope of the lines also increases with higher bond dimensions. Meaning, due to larger matrix and tensor sizes, the compute capabilities of GPU hardware are utilized in a more efficient manner. For $D=24576$, a performance of 11 TFLOPS has been reached with single GPU configuration, while for eight cards the same run resulted in 70 TFLOPS, which is very close to FP64 upper bound for eight NVIDIA A100-PCIE-40GB GPU devices ($P_{\rm max}=8\times 9.7$ TFLOP) ~\cite{nvidia}. The inset shows the scaling of the performance with respect to the estimated theoretical maximum as an inverse of the DMRG bond dimension.

\begin{figure}
  \centering
  \includegraphics[width=0.48\textwidth]{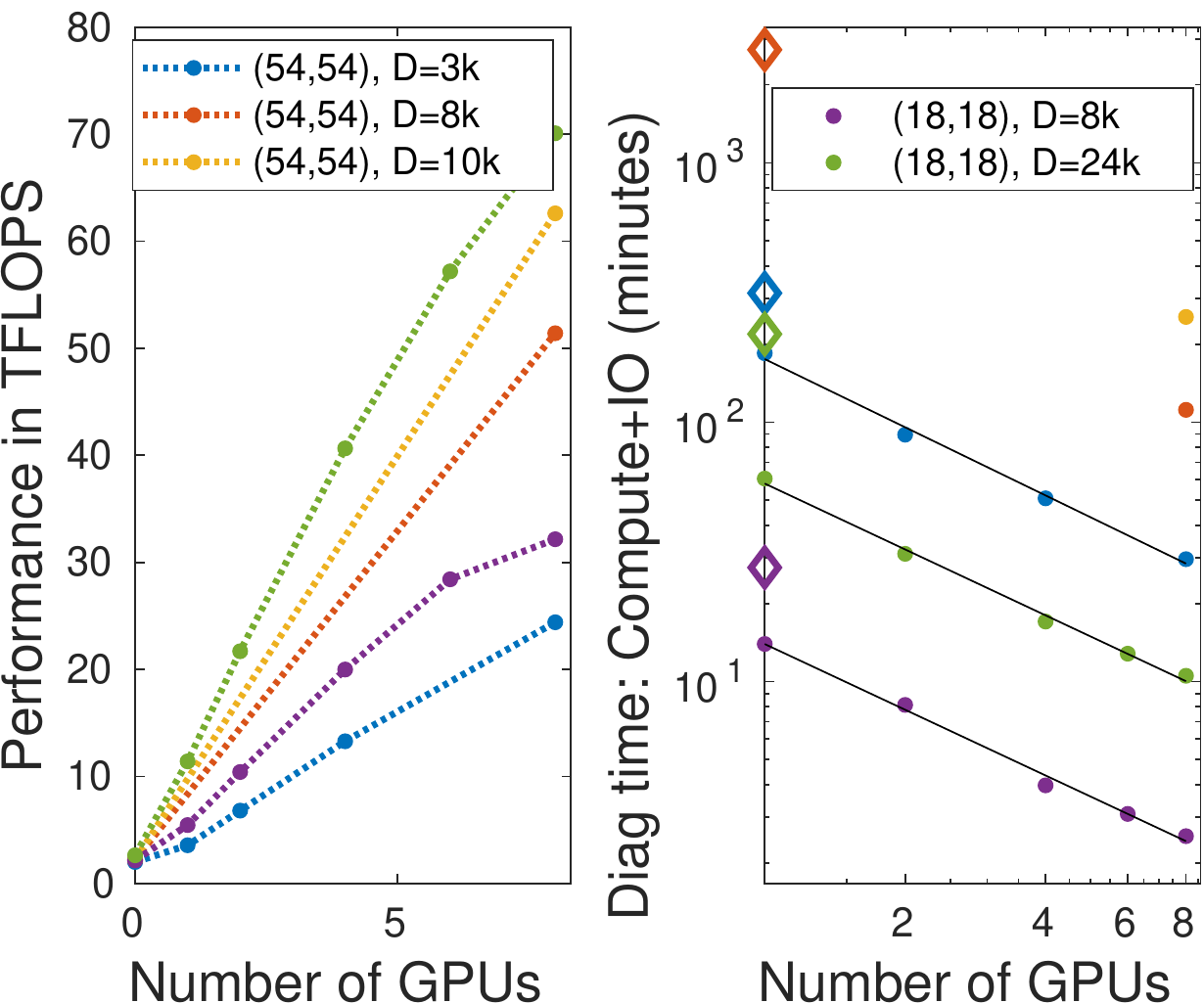}
  \caption{(Left panel)
  Performance measured in TFLOPS for two different CAS spaces as a function of the number of GPU devices for various fixed DMRG bond dimension values. CAS(18,18) and CAS(54,54) correspond to the F$_2$ and the FeMoco molecular systems, respectively. (Right panel) The total time in seconds used for the diagonalization of the effective Hamiltonian as a function of the number of GPU devices for the two molecular systems for fixed bond dimension. The open diamond symbols correspond to calculations performed using CPU only, i.e., when no GPU devices have been utilized. The dashed lines are guides to the eyes. For the FeMoco for $D=8192$ and $D=10240$ results only for 8 GPU devices are determined.
  The solid lines are results of
  first-order polynomial fits leading to exponents $-0.8438$, $-0.8416$ and $-0.8729$,
  respectively}
 \label{fig:perftime}
\end{figure}

The total execution time spent on the diagonalization procedure of the effective Hamiltionian
--- including both computational time related to the matrix-vector multiplications and the associated IO overhead ---
as a function of the number of GPU devices for the F$_2$ and FeMoco molecular systems for various fixed bond dimension values is summarized in Fig.~\ref{fig:perftime}. The open diamond symbols correspond to calculations performed using only CPU cores, i.e., when no GPU devices have been utilized. For very computationally demanding simulations (FeMoco) we present results only for eight GPU devices
obtained by dual AMD EPYC 7702 CPUs with $2\times64$ cores and eight NVIDIA A100-SXM4-40GB devices.
The dashed lines are guides to the eyes while
the solid lines are results of first-order polynomial fits.

\begin{figure}
  \centering
  \includegraphics[width=0.48\textwidth]{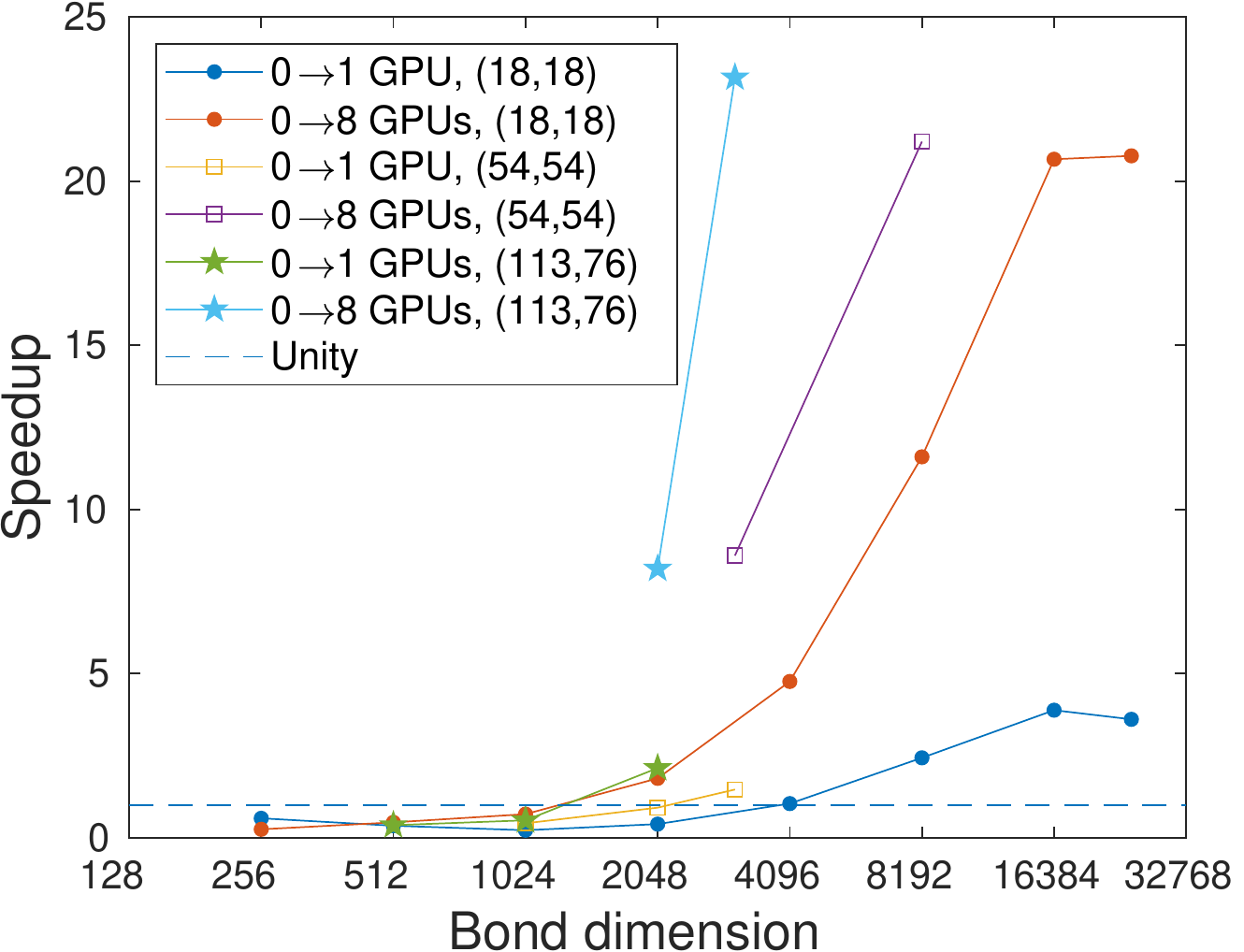}
  \caption{Speedup for selected data sets as a function of bond dimension. The dashed line at unity could be used to determine the minimal bond dimension for which the system dependent GPU accelerated solution becomes faster than the CPU only limit. 
}
 \label{fig:speedup-gpu}
\end{figure}

For a few selected data sets, the corresponding speedups measured with respect to the CPU-only limit is shown in Fig.~\ref{fig:speedup-gpu}. It is clearly visible that, using a single GPU device for intermediate and large $D$ values, a speedup by a factor of two to three can already be achieved. For small bond dimensions, however, the overhead might become dominant, thus as expected, calculations take more time in comparison to the CPU-only implementations. The system dependent minimal bond dimension for the crossover between the CPU-only and GPU accelerated calculations is indicated by the dashed line.

In addition to the speedup achieved by a single GPU, the measured total execution time versus the number of GPU devices falls on a line on a double logarithmic scale which holds regardless of the $D$ value and system size (see Fig.~\ref{fig:perftime} right panel). 
In fact, a first order polynomial fit leads to exponents 
$-0.8438$, $-0.8416$ and $-0.8729$, respectively. This means that doubling the number of GPU devices almost halves the total time and such perfect power-law scaling holds up to the eight GPU devices. 
The obtained exponents being close to minus one 
indicates that the IO communication overhead has been hidden efficiently behind the algebra
and the scaling is mainly determined by the computational complexity of the underlying matrix and tensor multiplications. 

For larger system sizes, i.e., for larger CAS spaces, larger TFLOP values are obtained and the speedup becomes even more remarkable compared to the CPU-only solution's limit. It is worth noting, that for smaller CAS(18,18) calculations a threshold has been reached for very large bond dimensions. This comes from the fact, that the jump from $D=16384$ to $D=24576$ yields a marginal $\approx 10$ percent increase in performance for both CPU-only (see Fig~\ref{fig:perf_cpu}) and multi-GPU (Fig.~\ref{fig:perf-gpu}) accelerated solutions. This is related to the fact, that, at first, with higher $D$ values the number of sectors increases together with the sector sizes, however for very large bond dimensions all sectors are populated and only sector sizes increase. For larger CAS spaces such saturation is expected only for much larger $D$ values.

\begin{figure}
  \centering
  \includegraphics[width=0.48\textwidth]{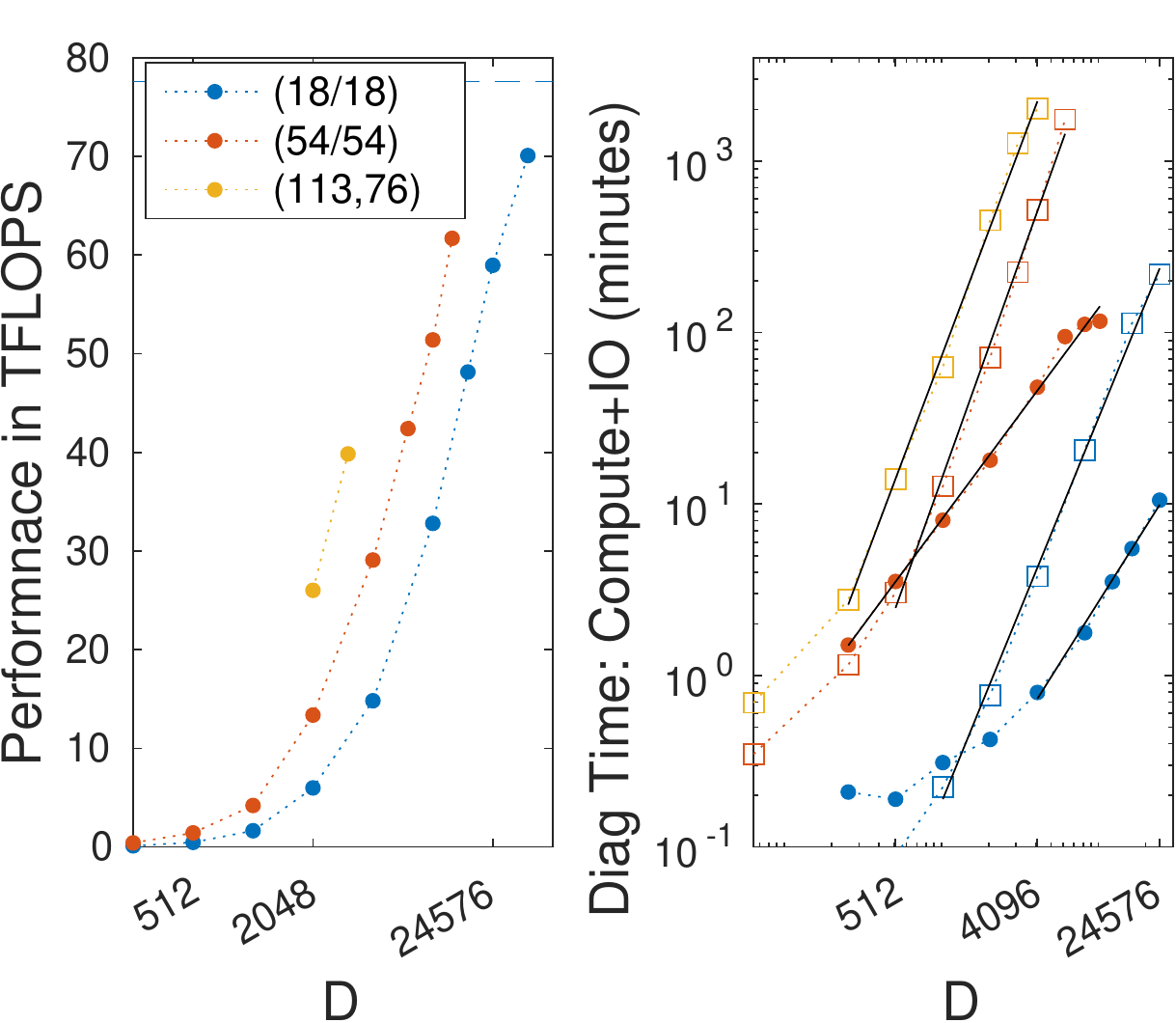}
  \caption{Performance and related total time for the eight GPU accelerated diagonalization procedure measured in TFLOPS and minutes, respectively, for the F$_2$ CAS(18,18) and FeMoco CAS(54,54) and CAS(113,76) as a function of DMRG bond dimension.
  In the right panel data corresponding to the CPU only limit is also included (square symbols).
  Dotted lines are guides to the eyes while the dashed line indicates the theoretical maximum. 
  The solid lines are results of first-order polynomial fits.
}
 \label{fig:perf-m}
\end{figure}

The performance and the related execution time via the diagonalization obtained with eight GPU devices are shown in Fig.~\ref{fig:perf-m} as a function of bond dimension for the F$_2$ and the FeMoco molecular systems. 
In the right panel data corresponding to the CPU only limit is also indicated by the square symbols. 
It is obvious from the right panel of the figure that data points fall on a line on a double logarithmic scale for an extended range of $D$ values. 
Here we emphasize that the scaling of the computational time for a single threaded calculation without utilization of symmetries based on quantum numbers is $D^3$. When quantum number based sector sparse representation is used together with our Maze-Runner solution (see Secs.~\ref{sec:methods} and ~\ref{sec:impl})
even for the CPU limit only we have obtained much smaller exponents being $2.24$, $2.53$ and $2.43$, respectively. Thus as expected, the utilization of symmetries together with CPU based parallelization provides a significant reduction in the scaling exponents which is further boosted by the GPU accelerated solution. 
For the 8 GPU accelerated solution a first order polynomial fit leads to exponents $1.45$ and $1.23$,
for CAS(18,18) and CAS(54,54), respectively.
The final exponents being close to one indicates that even a linear scaling has almost been reached.

To provide a better understanding on the related time scales, we have collected data from the right panel of Fig~\ref{fig:perftime} in Tab.~\ref{tab:exetime}. The matrix-vector multiplication time for our hybrid CPU-multiGPU solution, including the IO overhead, with F$_2$ CAS(18,18), $D=8129$ and three DMRG sweep takes 1.2 minutes. The same with $D=24576$ takes 10.3 minutes. These runs for the FeMoco CAS(54,54) with $D=3072$ and seven sweeps takes half an hour, while with a higher $D=8192$ it takes 1.9 hours. Lastly, with $D=10240$ the execution time goes up to 3.9 hours.
In addition, for the very large CAS(113,76) limit the execution time is still in the range of a few hours for seven sweeps and $D$ being in the range of a few thousands. 
This clearly demonstrates that GPU based technology can revolutionize the application of TNS based methods, for very large complex strongly correlated (multi-reference) problems. 

We have obtained similar results for various other quantum many body systems, like nuclear shell models \cite{Tichai-2022} and two-dimensional quantum lattice problems \cite{Zheng-2017,Krumnow-2021}, however, due to length restrictions these will be presented in a subsequent work.

\begin{table}[t]
  \centering
\begin{tabular}{l|c|c|r}
\hline
 \hline
 CAS & $D$ & \#of sweeps & Diag: Compute+IO\\
 \hline
 (18/18) & 8192  & 3 & 1.7 minutes\\
 (18/18) & 24756 & 3 & 10.3 minutes\\
 (54/54) & 3072  & 7 & 34 minutes \\
 (54/54) & 8192  & 7 & 1.9 hours\\
 (54/54) & 10240 & 7 & 3.9 hours\\
 (113/76) & 2048 & 7 & 1.6 hours\\ 
 (113/76) & 3072 & 7 & 2.2 hours\\
 \hline
 \hline
\end{tabular}
\caption{Total computation time together with IO overhead for the eight GPU accelerated diagonalization step for the F$_2$ and the FeMoco molecular systems for various DMRG parameters.
}
\label{tab:exetime}
\end{table}

\subsection{Renormalization}

In this section, we present our results obtained via the renormalization procedure, also know as the blocking step.
Unlike the diagonalization step, here several operators are constructed using series of tensor product operations and matrix multiplications. In addition, the memory demand of the underlying algebra also increases with the block size as all the position dependent operators withing the block must be generated. Therefore, much heavier IO overhead appears compared to the diagonalization step. This also requires slight modifications in the scheduling algorithm, but the same general framework outlined in Secs.~\ref{sec:methods} and ~\ref{sec:impl} has been applied.

The main bottleneck can be identified as the D2H CUDA kernels responsible for the retrieval of computed data on the devices, so they can be migrated into the final host-side data structures. On each device, before collection, the final matrices are the results of millions of operations defined by task tables. Such workflow is not so dissimilar to that of the diagonalization procedure. However, during renormalization, the sum reduction step is omitted, i.e. there is no summation over the position index within the block for matrix components of the individual quantum number dependent sectors.
What would have been just an addend, with no goal of ever escaping device memory, is now an end result needed to be transferred back to the host. This significantly enlarges the D2H IO kernel at the end of each renormalization call, making the procedure much more IO constrained.

\begin{figure}
  \centering
  \includegraphics[width=0.48\textwidth]{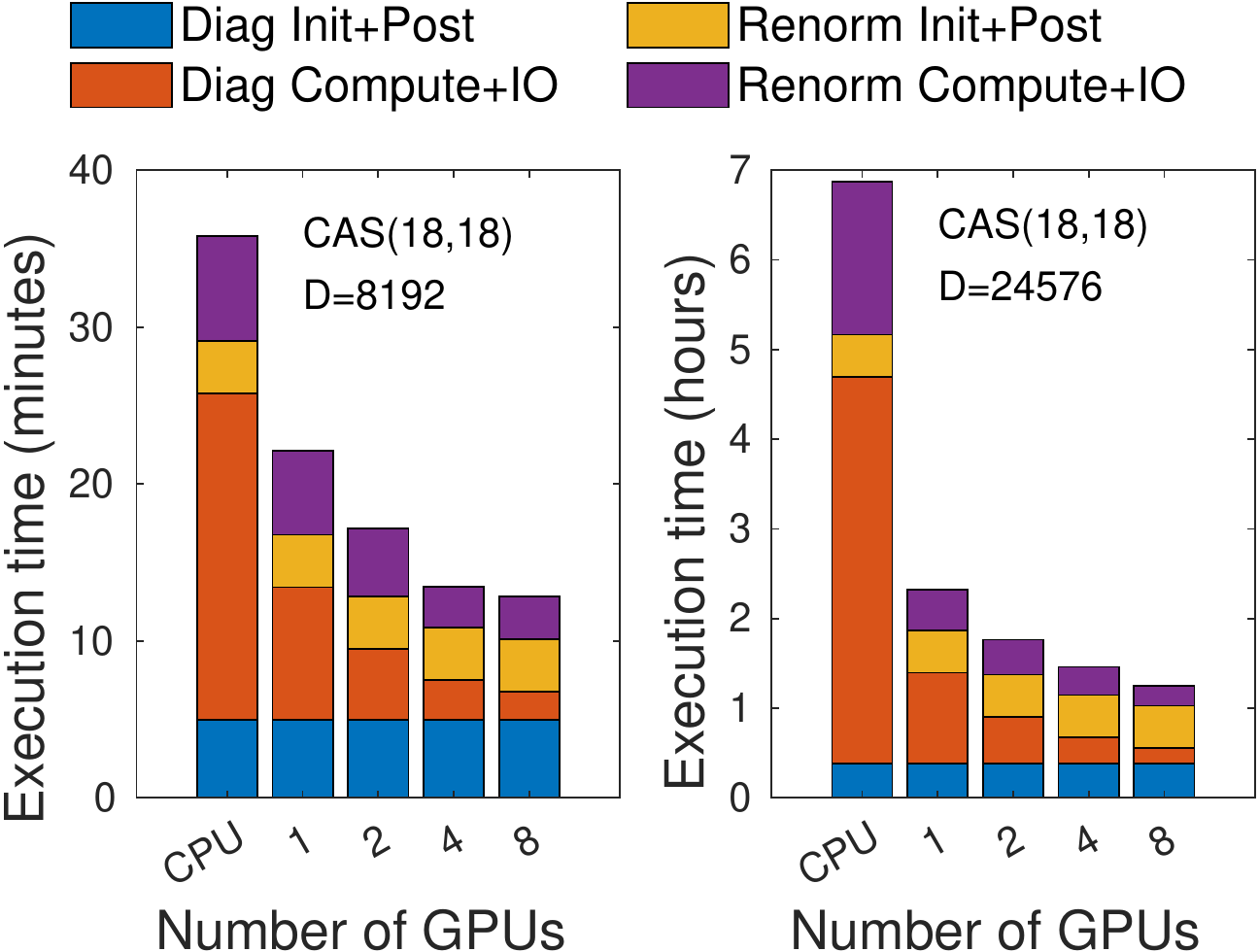}
  \caption{Decomposition of the total execution time according to the diagonalization and the renormalization steps for the F$_2$ molecule for CAS(18,18) orbital space using three DMRG sweeps and $D=8192$ (left panel) and for $D=24576$ (right panel).  
}
 \label{fig:renorm}
\end{figure}

The measured performance is summarized in Fig.~\ref{fig:renorm} as a function of the number of GPU devices for the F$_2$ CAS(18,18) molecule. Two different $D$ values were tested, each with 3 DMRG sweeps.

Here, the total execution time for the diagonalization and renormalization is split into two parts. In both cases, the CPU-only and GPU accelerated parts --- the latter together with its respective IO overhead --- are shown separately. Clearly, the execution time drops significantly for the renormalization procedure as well. Even with just a single device, the obtained speedup is remarkable. For the renormalization, however, a worse scaling is obtained compared to the diagonalization step due to the heavy IO demand leading to $\approx12$ TFLOPS of performance for 8 cards.

In fact, for $D=8129$, switching from 4 to 8 devices leads to a slight increase in execution time for the renormalization step. This is due to the compute time being already so small, that further parallelization yields only marginal improvements. From 4 GPUs onward, further increase to the number of devices will lead to the GPU overhead growing relatively faster, eventually overtaking the rate at which compute performance is gained from the extra devices. Thus, it not only cancels out the extra performance of the additional cards, but overall deteriorates the total execution time.

For larger $D$ values this is no longer the case, as higher computational complexity also means higher computational time during execution. This is what an increasing number of GPUs cuts down on, and with more to chop, there's more benefit regarding running times. For this reason, with higher $D$ values, the compute compares more favorably with both the IO and the GPU overhead.

While the penalty for using accelerators might slightly scale with $D$ values, even in worst case scenario, the overhead consists of IO operations tied to the underlying matrix and tensor algebra\footnote{In case it is not tied to the underlying algebra, then the overhead is unrelated to $D$ values, meaning it is in $\mathcal{O}(1)$ time complexity from the compute's point of view.}. This means, that every non-computational subroutine is in $\mathcal{O}(n^2)$ time complexity in regards to matrix leading dimensions, while the computation itself --- which benefits from continuously increasing GPU count --- is in $\Omega(n^3)$, as it contains large scale matrix multiplications.

FeMoco CAS(54,54) tells a similar story. Starting with small and intermediate $D\simeq3072$ values, the scaling is rather poor, but for larger $D$ value performance improves rapidly.

A natural step to improve the renormalization is to carry out the required algebra for the different operators on different nodes. Therefore, the most time consuming renormalization of a single operator will determine the overall execution time. Based on our measurements, we estimate that combining our hybrid CPU-multiGPU solution with our MPI implementation will lead to an immediate reduction of execution time for the renormalization procedure. We estimate an increase in performance by a factor of $10$.

In addition, the pre- and postprocessing steps related to preparing auxiliary operators via partial summations ~\cite{Szalay-2015a} can also be offloaded to accelerators, thus a large fraction of the time consumption shown by yellow color can be reduced and migrated into the GPU contribution shown by purple color. After such developments the yellow region will be almost invisible in Fig.~\ref{fig:renorm}.

Similarly, currently all algebra related to the L\'anczos and Davison diagonalization procedure, except the matrix vector multiplications, is performed on the CPU, but significant parts of these can also be migrated to accelerators. Therefore, the time demand indicated by blue color will become less than that of indicated by the red color. 

\subsection{Utilization of non-Abelian symmetries and more general tensor topologies}

From technical point of view, utilization of more symmetries increases the number of sectors for the quantum number based decomposed matrices and tensors which ultimately leads to lower memory demands, but more importantly to a significant increase in the number of independent tasks to be performed for the underlying algebra. For $U(1)$ symmetries this is a straightforward procedure, however, for non-Abelian symmetries a more delicate mathematical framework has to be utilized ~\cite{Mcculloch-2002,Toth-2008,Sharma-2012,Weichselbaum-2012,Keller-2016,Gunst-2019}. In our tensor library, the non-Abelian symmetry related layer is separated from the MPS layer~\cite{Werner-2020su2}, thus the hybrid CPU-multi-GPU kernel requires only minor modifications, i.e., each operations requires only an additional multiplication with a precalculated scalar and an extra data copy on the GPU devices. Therefore, some 20\% slowdown is expected, but usually a factor of 2 to 3 reduction in bond dimension can be achieved for the same accuracy which can  lead easily to an order of magnitude speedup. Details and results for our $SU(2)$ extended hybrid CPU-multiGPU kernel will be presented in a subsequent work.  

Our results have been demonstrated for the one-dimensional DMRG topology, but the hybrid CPU-multiGPU kernel can easily be extended for more general tensor network topologies ~\cite{Schollwock-2011,Szalay-2015a,Orus-2014}. When the number of subsystems (blocks) increases for more general networks, the number of the independent tasks of the underlying algebra also increases tremendously. This leads to significantly larger freedom to distribute data and tasks on significantly higher number of GPU devices, thus the GPU hardware provided computational power can be utilized even more efficiently.

It is also clear from Figs~\ref{fig:perf-gpu} and ~\ref{fig:perf-m} that, for large system sizes and bond dimensions, the optimal number of GPU devices has not been reached at eight cards. Therefore, an MPI based multiNode-multiGPU version is expected to further boost performance, introducing DMRG into the world of petascale computing~\cite{Ganahl-2023}. This will be analyzed in our subsequent work.  

\subsection{Massively parallel DMRG-RAS-X method}

Quite recently the DMRG algorithm has been cross-fertilized with the concept of restricted active space (RAS) method \cite{Barcza-2022b} and a rigorous mathematical analysis has lead to an efficient 
and stable extrapolation procedure to obtain the full-CI ground state energy within chemical accuracy using only limited CAS spaces~\cite{Friesecke-2022b}. Combination of this novel method, DMRG-RAS-X, with the hybrid CPU-multiGPU kernel (see results for FeMoco in Ref.~\cite{Friesecke-2022b}) has the potential to bring DMRG to a routinely applied method 
on the daily basis to complex strongly correlated (multi reference) problems requiring large orbital spaces. In addition, sampling the RAS space via the GPU accelerated solution on-the-fly together with additional mathematical developments will easily boost DMRG to handle efficiently static and dynamic correlations for systems with several hundreds of modes.

\subsection{Green DMRG}

Let us close our analysis by considering the power consumption of the TNS calculations, which nowadays are becoming one of the most important question due to high energy demands and costs. The thermal design power (TDP) for 2 $\times$ Intel(R) Xeon Gold 5318Y CPU is $2\times165$ Watts ~\cite{intel}, thus the estimated energy consumption for our benchmark results of 2.5 TFLOPS would lead to $\approx7.5$ GFLOPS/Watt.

In contrast to this, for an
NVIDIA A100-PCIE-40GB device the TDP is 250 Watts ~\cite{nvidia}. Therefore,
our 8 card accelerated hybrid algorithm with 70 TFLOPS performance results in
$70000/(330+8\times250)=30.04$ GFLOPS/Watt. This means power efficiency could be increased by a factor of four. In other words, for a given calculation the cost of the energy demand arising from the processors can be reduced to one quarter of the original consumption. Here, we simply took the upper bounds of the related quantities while in practice the energy consumption of the GPU devices fluctuates significantly during the calculations, thus even a better ratio can be obtained.

\section{Conclusion}
\label{sec:conclusion}

\begin{figure}
  \centering
  \includegraphics[width=0.48\textwidth]{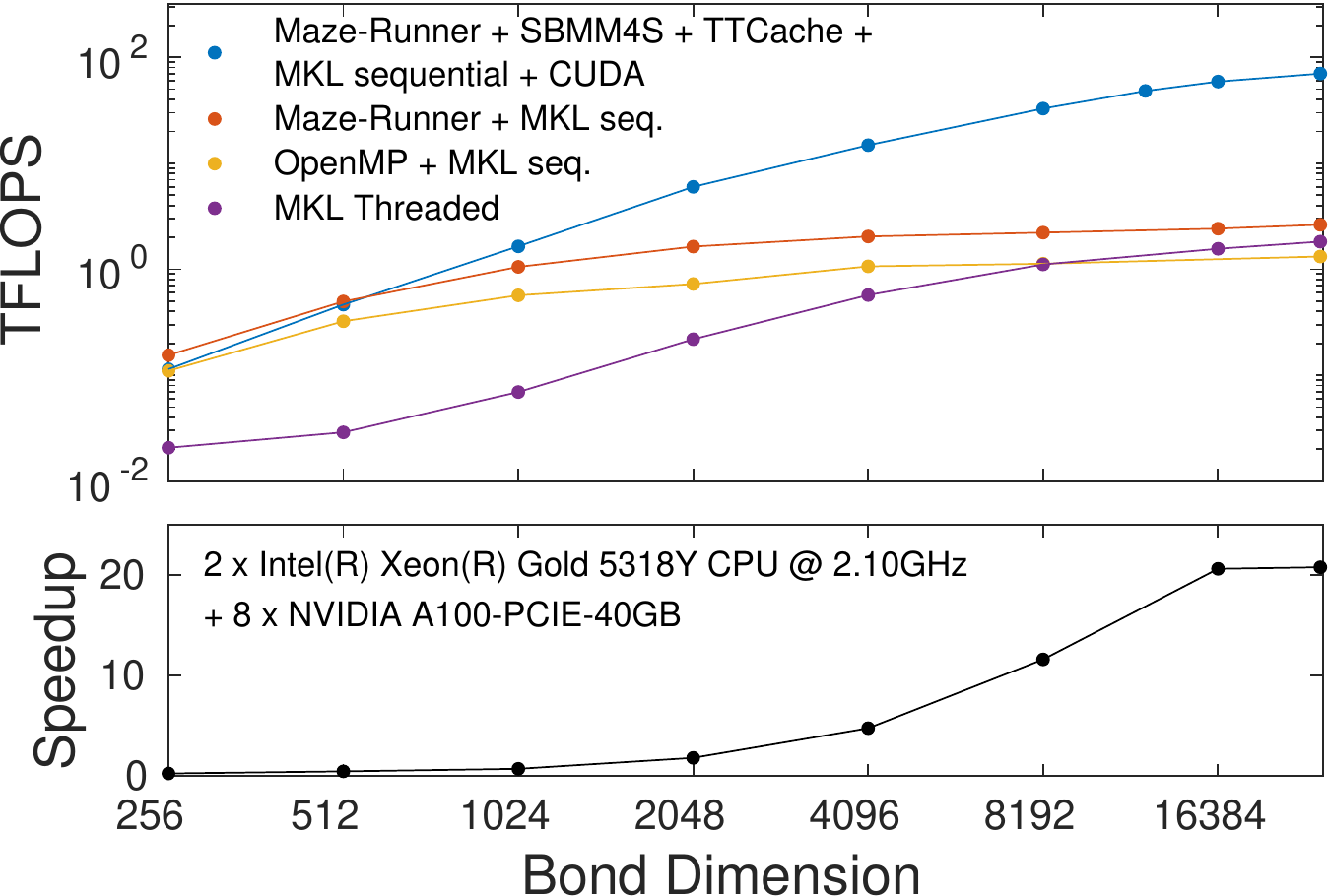}
  \caption{Data recollected from Figs.~\ref{fig:perf_cpu} and~\ref{fig:speedup-gpu} to summarize final performance achieved via the diagonalization procedure for the F$_2$ molecule in CAS(18,18) orbital space.
  Speedup is measured between the $8\times$GPU accelerated solution (blue) and the CPU-only version (orange).}
  \label{fig:algo_war}
\end{figure}

In this work, we have presented novel algorithmic solutions together with implementation details to extend current limits of tensor network state algorithms on high performance computing infrastructure building on state-of-the-art hardware and software technologies. These includes the following main contributions (for related performance measures see Fig.~\ref{fig:algo_war}):
\begin{itemize}
  \item \textbf{Maze-Runner:} Lightweight, low complexity parallel construct enabling producer-consumer like task handling, but without the need to enforce roles on threads. Due to homogeneous threading and task creation prioritisation, scheduling is simplified to the point of triviality. This leads to low overhead and lack of scheduling imperfections when confronted with volatile task creation / execution times. 
  \item \textbf{TTCache:} Virtual memory addressing designed to vastly reduce redundant IO operations and eliminate memory fragmentation as well as allocation overhead. TTCache works by factorizing data into attributes, then hierarchically mapping such attributes to execution blocks. Execution is done by traversing a tree-like structure, in which nodes close to each other depend on largely the same set of attributes.
  \item \textbf{SBMM4S:} Batched type matrix multiplication with inherent zero cost sum reduction. Produces a single result by multiplying an entire batch of matrices with concatenated vector arrays of interleaved matrices. Intermediate results of chained matrix multiplications are reached using strided batched type matrix multiplications with specific offset values to enable interleaving.
\end{itemize}
Benchmark results have been presented for the F$_2$ molecule, that is the limit of exact diagonalization on super computers, and for the FeMoco cluster that is in the focus of modern quantum chemistry and also subject of method development benchmarks. We have demonstrated that our massively parallel hybrid CPU-multiGPU architecture can perform the diagonalization of the effective quantum many body Hamiltionian
very efficiently by utilizing the full capacity of modern CPU and GPU hardware. 

When GPU devices are in play, even with just single GPU configurations, a speedup by a factor of two to four has been achieved for larger $D$ values. In addition, based on our measurements, we found a linear relation between performance and the number of used accelerators. With such efficient scaling, the theoretical upper bound of 77.6 TFLOPS for eight NVIDIA A100-PCIE-40GB GPU devices has almost been reached. Our efforts pay well in reduction of computational time as the total time as a function of the number of GPU devices drops linearly on a doubly logarithmic scale. Therefore, utilizing eight GPU devices leads to over 20 times the performance when $D$ values are large enough.

As part of our current work, we are further optimizing the $SU(2)$ spin adapted GPU solution for intermediate bond dimension values, developing and implementing new scheduling systems and optimization protocols, converting other parts of the DMRG method to highly efficient GPU accelerated code and migrating our MPI protocols to the newly developed hybrid CPU-multiGPU solution, so that multiple nodes can be utilized. These upgrades together could easily extend the performance of the DMRG method to petascale computing for interacting quantum many body problems with long range interactions.   

%------------------------------------------------

\section*{Acknowledgement}
The authors acknowledge useful discussions with Tam\'as Kozsik and Mikl\'os Werner.
This research was supported 
by the Hungarian National Research, Development and Innovation Office (NKFIH) through Grant Nos.~K134983 and TKP2021-NVA-04
by the Quantum Information National Laboratory of Hungary, 
and by the Hans Fischer Senior Fellowship programme funded by the Technical University
of Munich – Institute for Advanced Study. 
\"O.L. has also been supported
by the Center for Scalable and Predictive methods
for Excitation and Correlated phenomena (SPEC),
funded as part of the Computational Chemical Sciences Program by the U.S. Department of Energy
(DOE), Office of Science, Office of Basic Energy Sciences, Division of Chemical Sciences, Geosciences, and Biosciences at Pacific Northwest National Laboratory.
We thank computational support from the Wigner Scientific Computing Laboratory (WSCLAB), 
the Eötvös Loránd University
and the Governmental Information-Technology Development Agency providing access to the supercomputer Komondor.
The simulations have also been performed on the national supercomputer HPE Apollo Hawk at the High Performance Computing Center Stuttgart (HLRS) under the grant number MPTNS/44246.

%----------------------------------------------------------------------------------------
%	 REFERENCES
%----------------------------------------------------------------------------------------

%----------------------------------------------------------------------------------------

\end{document}